\documentstyle[11pt,equations,doublespace,cite]{article}
\def\ksl{\hbox{\hbox{${k}$}}\kern-1.9mm{\hbox{${/}$}}}
\def\qsl{\hbox{\hbox{${q}$}}\kern-1.9mm{\hbox{${/}$}}}
\def\pisl{\hbox{\hbox{${p}$}}\kern-1.7mm{\hbox{${/}$}}}
\newfont{\bbiga}{cmti10 scaled \magstep2}
\newcommand{\smallz}{{\scriptscriptstyle Z}} 
\newcommand{\smallw}{{\scriptscriptstyle W}} %
\newcommand{\LT}{{\scriptscriptstyle L.T.}} %
\newcommand{\mz}{m_\smallz}
\newcommand{\mw}{m_\smallw}
\newcommand{\bibar}{\mbox{$\bar{b}$}}
\newcommand{\qbar}{\bar{q}}
\newcommand{\pib}{\mbox{$\Pi_b$}}
\newcommand{\pifi}{\mbox{$\Pi_\Phi^{{\scriptscriptstyle (1)}}$}}
\newcommand{\pitwo}[1]{\mbox{$\Pi^{{\scriptscriptstyle (2)}}_{{\rm #1}}$}}
\newcommand{\pig}{\mbox{$\Pi_g^{{\scriptscriptstyle (1)}}$}}
\newcommand{\piov}[1]{\mbox{$\overline{\Pi}_{
           {\scriptscriptstyle 6({\rm b}_i)}}^{#1}(p,p^\prime,\qbar)$}}
\newcommand{\piovw}[2]{
        \mbox{$\overline{\Pi}_{
        {\scriptscriptstyle 6({\rm b}_{#1})}}^{#2}$}}
\newcommand{\alfa}{\mbox{$ \alpha \frac{m_t^2}{\mw^2} $}}
\newcommand{\alfas}{\mbox{$ \alpha \alpha_s \frac{m_t^2}{\mw^2} $}}
\newcommand{\selfs}{self-energies}
\newcommand{\self}{self-energy}
\newcommand{\bre}{bremsstrahlung}
\newcommand{\cor}{corrections}
\newcommand{\ew}{electroweak}

\newcommand{\equ}[1]{Eq.(\ref{#1})}
\newcommand{\eqs}[2]{Eqs.(\ref{#1},\ref{#2})}
\newcommand{\efe}[1]{Ref.\cite{#1}}
\newcommand{\efs}[2]{Ref.\cite{#1,#2}}

\newcommand{\cvi}[1]{ {\cal V}_{#1}^\mu }
\newcommand{\cuu}{ {\cal U}_{\Phi}^\mu }
\newcommand{\cud}{ {\cal U}_{\Phi g}^\mu }
\newcommand{\cutu}{\widetilde{{\cal U}}_{\Phi}^\mu }
\newcommand{\cutd}{\widetilde{{\cal U}}_{\Phi g}^\mu }
\newcommand{\dvu}{ {\cal D}_{\Phi}}
\newcommand{\dvd}{ {\cal D}_{\Phi g}}
\newcommand{\be}{\begin{equation}}
\newcommand{\ee}{\end{equation}}
\newcommand{\een}{\end{subequations}}
\newcommand{\ben}{\begin{subequations}}
\newcommand{\beq}{\begin{eqalignno}}
\newcommand{\eeq}{\end{eqalignno}}
\newcommand{\bea}{\begin{eqnarray}}
\newcommand{\eea}{\end{eqnarray}}
\newenvironment{appendletter}
 {
  \typeout{ Starting Appendix \thesection }
  \setcounter{equation}{0}
  
 }{
  \typeout{Appendix done}
 }
\renewcommand{\thefootnote}{\fnsymbol{footnote} }
\begin{document}
\begin{titlepage}
\begin{flushright}
            DFPD 93/TH/03 \\
            January 1993
\end{flushright}
\begin{center}
{\Large Current Algebra Approach to Heavy Top Effects in $Z \rightarrow
b+ \bibar$}\\
\vspace{1.5cm}
Giuseppe Degrassi\\
{\em Department of Physics, New York University, New York, NY 10003,
U.S.A.\\
and \\
Dipartimento di Fisica, Universit\`a  di Padova,
Sezione INFN di Padova \\
Via Marzolo 8, 35131 Padova, Italy \footnote{Permanent address}}\\
\vspace{1cm}
\vspace{1cm} ABSTRACT
\end{center}
\vspace{0.5cm}
The $O(\alfa)$ and $O(\alfas)$ \cor\ to the partial decay width $\Gamma(Z
\rightarrow b + \bibar)$
are computed using the current algebra formulation of
radiative \cor. This framework allows one to easily enforce the relevant
Ward identity that greatly simplifies the calculations. As a result,
the one-loop
$O(\alfa)$ contribution is computed through the investigation of only two
convergent diagrams. The computation of the QCD \cor\ to the one-loop
$O(\alfa)$ term involves fewer  diagrams than in the standard approach.
In particular, the number of infrared divergent contribution is reduced. The
calculation is performed in the dimensional regularization scheme and no term
more divergent than $\frac{1}{n-4}$ is found. Our result confirms
the screening of the one-loop top mass effect recently found by Fleischer
{\em et al.}.
\end{titlepage}
\setcounter{footnote}{0}
\renewcommand{\thefootnote}{\arabic{footnote}}
\section{Introduction}
It is well known that the \ew\ radiative parameters $\delta \rho$ \cite{b1}
and $\Delta r$ \cite{b2} are affected by virtual top exchanges in the
\selfs\ of the vector bosons through terms that depend quadratically on
$m_t$, the mass of the top quark. Moreover, due to the isodoublet nature of the
top, $m_t^2$ \cor\ are  also found in the
$Z b \bibar$-vertex.
\par The status of the theoretical calculations of these \cor\ is quite
advanced. At the moment we have available for the vacuum polarization
functions, beside the one-loop calculation \cite{b3bis}, the complete
perturbative $O(\alpha \alpha_s)$ QCD contribution \cite{b4,b5}, studies
of the $t \bar{t}$ threshold effects \footnote{For an updated analysis
of the QCD effects in the \ew\ \cor\ see Ref.\cite{b5bis} and references
therein.} and the leading $O( ( \alfa)^2 )$ term \cite{b6,b7}. Concerning the
$Z b \bibar$-vertex, the one-loop calculation was performed by several groups
a few years ago \cite{b8,b9,b10}. Recently, the leading $ O(\alfas)$ \cite{b11}
and the $O((\alfa)^2)$ terms \cite{b7} have been computed.
\par The calculation of the two-loop corrections has proven to be a difficult
task. Indeed, apart from the perturbative QCD contribution to the vacuum
polarization,  the two-loop contributions have been evaluated in some
approximation, either neglecting all the masses and momenta but the top
mass \cite{b6,b11} or retaining besides the top only the higgs mass \cite{b7}.
However, these approximations are sufficient to derive the most interesting
part of the corrections, namely the leading $m_t^2$ contribution.
\par It is the aim of this paper to apply techniques used in the current
algebra formulation of radiative \cor\ \cite{b12} to derive the leading
$O(\alfa)$ and $O(\alfas)$ corrections in the partial decay width
$\Gamma(Z \rightarrow b+ \bibar)$.
As will be seen in the following, these techniques
greatly simplify the calculations. In the most interesting case, namely the
two-loop $O(\alfas)$ correction, in addition to facing  fewer
contributions than in the standard calculation \cite{b11} one deals with
less divergent diagrams . Furthermore, the treatment of the infrared divergent
(IR) contribution is greatly simplified. Our result shows that the QCD
\cor\ have opposite sign to that of the one-loop contribution, leading
then an increase in the $m_t$ upper bound. We find agreement with the recent
calculation of the $\alpha_s G_\mu\,  m_t^2$ term by Fleischer {\em et al.}.
\par The paper is organized as follows. In Section 2 the current algebra
derivation of the $O(\alfa)$ one-loop term is
presented. Section 3 is devoted to the calculation of the QCD \cor\ to
this term. In Section 4 we discuss the results. The Appendices give some
details about the QCD calculation. In Appendix A we show that the formulae
 used for the four and five-point correlation
functions (Cfr. \eqs{e19a}{e17}) contain correctly the two-loop wave function
renormalization of the external legs. Appendix B lists the individual
contributions of the various diagrams.
\section{One-loop $O(\alfa)$ correction}
In this section we derive the leading one-loop $O(\alfa)$ term in the
$Z b \bibar$-vertex using two and three-point correlation
functions. This allows us to set the framework for our subsequent
discussion of the QCD \cor\ and to derive the basic Ward identity that
enters into the calculation.
\par In order to fix our notation we write the part of the Standard Model
(SM) Lagrangian density that describes the interaction of the $W,\,Z$
and unphysical scalars with fermions as \cite{b12}
\be
{\cal L}_{int}  = - \frac{g}{\sqrt{2}}(W^{\dagger}_\mu J_W^\mu +  h.c.)
        -  \frac{g}{c}Z_\mu J^\mu _Z
        - \frac{g}{2\, \mw} \left[ \Phi_2 S_2 + \sqrt2 \,( \Phi^\dagger
          S + h.c.) \right] \label{e1}
\ee
where $g$ is the $SU(2)$ coupling, $\mw$ stands for the mass of the $W$ boson,
$c$ is an abbreviation for $\cos \theta_W$, $J_Z^\mu$ and $J_W^\mu$ are the
fermionic currents coupled to $Z$ and $W$ respectively,  $W^\dagger$ is the
field that creates a $W^+$ meson, $\Phi_2$ and $\Phi$ the unphysical
counterparts associated with the $Z$ and $W$ and \ben \label{e2} \beq
S_2 =& 2 \,\partial_\mu J^\mu_Z = -i \bar{\psi}\, m^0\, C_3 \gamma_5 \psi
                   \label{e2a} \\
S =& -i \, \partial_\mu J^\mu_W = \bar{\psi} \Gamma \psi. \label{e2b}
\eeq
In \eqs{e2a}{e2b} $\psi$ represents the column vector $\psi \equiv (t,b)^T, \:
m^0, \: C_3$ and $\Gamma$ are the $2 \times 2$ matrices \beq
m^0 =& \left(  \begin{array}{cc} m_t^0 & 0 \\
                                  0 & 0 \end{array} \right) \label{e2c} \\
C_3 =& \left(  \begin{array}{cr} 1 & 0 \\
                                 0 & -1 \end{array} \right) \label{e2d} \\
\Gamma =& \left(  \begin{array}{cc} 0 & 0 \\
                   -m_t^0 \, a_+ & 0 \end{array} \right) , \label{e2e}
\eeq \een
$a_+ \equiv \frac{1+\gamma_5}{2}$ and the superscript
$0$ on $m_t$ refers to the bare mass.
As it is evident from \eqs{e2c}{e2e} we are considering only the third
generation and taking the bottom quark as massless.
\par The one-loop vertex diagrams contributing to the $O(\alfa)$
correction are depicted symbolically in Fig.1. The circles represent the sum
of diagrams in which the ends of the $\Phi^{\pm}$ propagators are attached in
all possible ways to the external $ b\bibar$ quarks.
\par The amplitude corresponding to Fig.1(a) can be expressed as \cite{b12}
\ben \label{e3a} \beq
 \cvi{\Phi} =&   -i \,\frac{g}{c} \, \lim_{\qbar \rightarrow q}\,
                T^\mu_\Phi (\qbar, \, p,\, p^\prime) \label{e3}
\eeq
with \beq
T^\mu_\Phi =& \frac{-i}{2} \frac{g^2}{2 \,\mw^2}
\int \frac{d^n k}{(2 \pi)^n\, \mu^{n-4}} \frac1{k^2 - \mw^2}
\int d^n y \,e^{-i \qbar \cdot y} \nonumber \\
\times & \int d^n x \,e^{ik \cdot x} < p^\prime p \, | \, T^*
\left[ J_Z^\mu (y)\,
( S^{\dagger}(x) \, S(0)+ h.c.) \right] \, | \,0 >  ,\label{e4}
\eeq \een
where $n$ is the dimension of  space--time, $\mu$ is the 't Hooft mass
scale, $T^*$ is the covariant time--ordered product, $p$ and $p^\prime$ the
momenta of the $\bibar$ and $b$ quark respectively and $q= p+p^\prime$ the
momentum carried by the $Z$. As the $O(\alfa)$ term should be gauge invariant
by itself we find it convenient to carry out the calculation in the
't Hooft--Feynman gauge in which the propagator of the $\Phi$ field has
the form $i(k^2 - \mw^2)^{-1}$.\\
As shown in \efs{b13}{b12}, the limiting procedure in \equ{e3}\ affects the
insertion of the $\Phi$ fields on the external lines in such a way that
$\cvi{\Phi}$ contains not only the proper vertex correction but also the
contributions from the wave function renormalization of the $b$ quarks.
\par To trigger a Ward identity we contract $T^\alpha_\Phi$ with $\qbar_\alpha$
obtaining \be
\qbar_\alpha T^\alpha_\Phi = \dvu +
\int \frac{d^n k}{(2 \pi)^n\, \mu^{n-4}} \frac1{k^2 - \mw^2} \left[
V_\Phi (k - \qbar) - V_\Phi (-k-q) \right] \label{e5}
\ee
where $\dvu$ is an expression analogous to $T^\mu_\Phi$ with the
replacement $J^\mu_Z \rightarrow -i \partial_\mu J^\mu_Z$ and \be
V_\Phi (k) = - \frac{g^2}{ 4 \,\mw^2} \frac{c^2- s^2}{2}
\int d^n x \,e^{ik \cdot x} < p^\prime p \, | \, T^*
( S^{\dagger}(x) \, S(0) - S^\dagger (0) S(x)) \, | \,0 >  ,\label{e6}
\ee
with $s^2 = 1- c^2$. In deriving \equ{e6}\ we have used the commutation
relation \be
\delta (x_0 - y_0) \left[ J_Z^0 (x), \, S (y) \right]
        = - \frac{c^2 - s^2}2 S (x) \delta^n (x-y). \label{e7b}
\ee
Differentiating \equ{e5}\ with respect to $\qbar_\mu$ one obtains \be
T^\mu_{\Phi} = - \qbar_\alpha \frac{\partial}{\partial \qbar_\mu} T^\alpha_\Phi
         + \frac{\partial}{\partial \qbar_\mu} \dvu -
\int \frac{d^n k}{(2 \pi)^n\, \mu^{n-4}} \frac1{k^2 - \mw^2}
\frac{\partial}{\partial k_\mu} V_\Phi (k - \qbar). \label{e7}
\ee
It is easy to show following, for example,
 the discussion in Section VII of \efe{b12}
that the first term in the r.h.s.\ of \equ{e7}\ is $O(\alpha)$ instead of
$O(\alfa)$. Terms of this kind will be considered throughout
all our calculations
as subleading and therefore neglected. After a partial integration in the
last term  in  \equ{e7}\ we can write for $T^\mu_\Phi$ \beq
T^\mu_{\Phi} =&  \frac{\partial}{\partial \qbar_\mu} \dvu +
                                 \frac{g^2}{4\, \mw^2}
(c^2- s^2) \int \frac{d^n k}{(2 \pi)^n\, \mu^{n-4}} \frac{k^\mu}{(k^2 -
\mw^2)^2}
\nonumber \\
\times & \int d^n x \,e^{i(k-\qbar) \cdot x} < p^\prime p \, | \, T^*
( S^{\dagger}(x) \, S(0) - S^\dagger (0) S(x)) \, | \,0 > .\label{e8}
\eeq
As we are interested in the leading $m_t$ dependent contribution, we can
set $\qbar=q=0$ in the second term of \equ{e8}\ and obtain
\beq
(T^\mu_{\Phi})_{\LT} = & \frac{\partial}{\partial \qbar_\mu} \dvu +
\frac{g^2}{2\, \mw^2}
(c^2- s^2) \int \frac{d^n k}{(2 \pi)^n\, \mu^{n-4}}
\frac{k^\mu}{(k^2 - \mw^2)^2} \nonumber \\
\times &\int d^n x \,e^{ik \cdot x} < p^\prime p \, | \, T^*
( S^{\dagger}(x) \, S(0))  \, | \,0 > ,\label{e9}
\eeq
where the subscript $L.T.$ reminds us that we are considering only the
leading term.
\par Turning our attention to Fig.1(b) we write the amplitude as
\ben \label{e10} \beq
\cuu =& -i \frac{g}{c} \cutu \label{e10a}\\
\cutu =& - \frac{g^2 }{2\,\mw^2} (c^2- s^2)
\int \frac{d^n k}{(2 \pi)^n \, \mu^{n-4}} \frac{k^\mu}{(k^2-\mw^2)
[(k+q)^2 -\mw^2]}  \nonumber \\
& \times \int d^n x \,e^{ik \cdot x}
< p^\prime p \, | \, T^*  (S^\dagger (x)S(0))\, | \,0 >  .\label{e10b}
\eeq \een
It is immediate to see that the leading term in $\cutu$, obtained by putting
$q=0$ in \equ{e10b}, cancels exactly the second term in the r.h.s.\ of
\equ{e9}. We conclude  that the leading $O(\alfa)$ correction in the
$Z b \bibar$-vertex is given by $ \left. -i \frac{g}{c}
\frac{\partial}{\partial \qbar_\mu} \dvu \right|_{\qbar = q}$ or, recalling
\equ{e2a},  \beq
(\cvi{\Phi} + \cuu )_{\LT} =& i \frac{g^3}{8 c\, \mw^2}
\frac{\partial}{\partial \qbar_\mu} \left\{
\int \frac{d^n k}{(2 \pi)^n\, \mu^{n-4}} \frac1{k^2 - \mw^2}
\int d^n y \,e^{-i \qbar \cdot y} \right. \nonumber \\
& \left. \left.
\times  \int d^n x \,e^{ik \cdot x} < p^\prime p \, | \, T^*
\left[ S_2 (y)\,
( S^{\dagger}(x) \, S(0)+ h.c.) \right] \, | \,0 > \right\}
\right|_{\qbar = q}  .\label{e11}
\eeq
\par To express the r.h.s.\ of \equ{e11}\ in terms of Feynman diagrams we
use Wick's theorem. We now observe that a non-zero contribution is obtained
only when $S_2$ is acting on a top line. Fig.2 represents schematically
the fermion lines in \equ{e11}\ for $\qbar \neq q$ before the $k$ integration
is performed. The dotted line stands for the $\qbar$ momentum absorbed by the
$S_2$ operator while the dashed lines indicate the momenta emitted or
absorbed by the
$S$ and $S^\dagger$ operators. For example, in Fig.2(a) $\qbar$ is absorbed
by $S_2(y)$ while $k$ is emitted by $S^\dagger (x)$. Similarly in Fig.2(b)
$k$ is emitted by $S(x)$ and $\qbar$ is absorbed by $S_2(y)$.
\par Using an anticommuting $\gamma_5$ the contributions of Fig.2(a) and
Fig.2(b) to \equ{e11} are
\ben \label{e12} \beq
2({\rm a}) =&
\frac{g^3}{8c} \frac{m_t^4}{\mw^2} \frac{\partial}{\partial \qbar_\mu}
\left. \left\{ \int \frac{d^4 k}{(2 \pi)^4} \frac{\bar{u}(p^\prime) \bar{\qsl}
a_- v(p)}{[k^2 - \mw^2]\,[(k-p+\qbar)^2 - m_t^2]\,[(k-p)^2- m_t^2]}
\right\} \right|_{\qbar=q} \label{e12a} \\
2( {\rm b}) =&
 \frac{g^3}{8c} \frac{m_t^4}{\mw^2} \frac{\partial}{\partial \qbar_\mu}
\left. \left\{ \int \frac{d^4 k}{(2 \pi)^4} \frac{\bar{u}(p^\prime) \bar{\qsl}
a_- v(p)}{[k^2 - \mw^2]\,[(k+p^\prime-\qbar)^2 - m_t^2]\,[(k+p^\prime)^2-
m_t^2]} \right\} \right|_{\qbar=q} \label{e12b}
\eeq \een
where the l.h.s.\  in the above equations indicates the appropriate diagram
in Fig.2 and $a_- \equiv \frac{1-\gamma_5}{2}$.
It is easy to see that the differentiation with respect to $\qbar_\mu$
of the denominators in Eqs.(\ref{e12}) gives a zero contribution after the
limit $\qbar \rightarrow q$ is taken. Differentiating  the numerators and then
putting $p^\prime=p=\mw=0$ one obtains \be
(\cvi{\Phi} + \cuu )_{\LT} = -i \frac{g}{2c} \frac{g^2}{16 \pi^2}
\frac{m_t^2}{2\,\mw^2} \bar{u}(p^\prime) \gamma^\mu a_- v(p). \label{e13}
\ee
\equ{e13} gives the contribution of the leading $O(\alfa)$ vertex correction.
In the
on--shell scheme where $s^2 \equiv \sin^2 \theta_W $ is defined according to
\cite{b2}
$$ \sin^2 \theta_W =1 - \frac{\mw^2}{ \mz^2} $$
and the partial width $\Gamma (Z \rightarrow f + \bar{f})$, where $f$ is a
massless fermion,  can be written as \cite{b14} \be
\Gamma( Z \rightarrow f + \bar{f})
= N_c^f \frac{ G_\mu}{\sqrt{2}} \frac{\mz^3}{1-\delta \rho_{irr}}
      \frac{| \rho_{ff}(\mz^2)| }{12 \pi}
\left\{ 1 - 4 I_f^3 Q_f s^2 Re\, k_f + 8 Q_f^2 s^4 |k_f|^2 \right\}
 \label{e14}\\
\ee
we see that the term (\ref{e13}) introduces a
modification in the \ew\ form factors
$\rho_{bb}$ and $k_b$
equal to $- 4 x_t$ and $+2x_t$ respectively, where
$x_t = (G_\mu\, m_t^2)/(\sqrt2\, 8 \pi^2)$.
In \equ{e14}\ $N_c$ refers to the color factor, $I^3_f$ and $Q_f$ stand
for the $I_3$ and charge quantum number of the fermion and
$\delta \rho_{irr} = 3 x_t$.
The counterpart
of $\rho_{bb}$ and $k_b$ in the $\overline{MS}$ scheme, namely the
quantity $\overline{\rho}_{bb}$ and $\hat{k}_b$ \cite{b14}, will be
accordingly modified by the factors $- \hat{\alpha}/(4 \pi\, \hat{s}^2)
(m_t^2/ \mw^2)$ and $\hat{\alpha}/(4 \pi\, \hat{s}^2)
m_t^2/(2\, \mw^2)$, where $\hat{\alpha}= \hat{e}^2(\mz)/(4 \pi)$ and
$\hat{s}^2 \equiv \sin^2 \hat{\theta}_W (\mz)$ are the $\overline{MS}$ coupling
and weak angle evaluated at $\mu=\mz$.
\section{Two-loop $O(\alfas)$ correction}
We proceed to compute the QCD \cor\ to the leading $O(\alfa)$ term, i.e.\
the $O(\alfas)$ contribution. We begin by noticing that the wave function
renormalization of the external $Z$ has no contribution $O(\alfas)$, therefore
all the QCD \cor\ are obtained by adding a gluon, in all possible ways, to the
diagrams of Fig.1. The result is shown in Fig.3. The circles represent now
the sum of diagrams of various kind. i) Diagrams where both the ends of the
$\Phi$ as well as the gluon propagator are attached to the external $b, \,
\bibar$ lines. These are reducible diagrams contributing to the wave function
renormalization of the external fermions. ii) Diagrams where either the $\Phi$
or the gluon propagator has both ends attached to an external line while
the other propagator is acting internally. iii) Diagrams where each propagator
has one end attached to an external fermion while the other end is attached
internally.
\par Fig.3(a) can be expressed as a five-point correlation function or
\ben \label{e16} \beq
 \cvi{\Phi g}& =  -i \,\frac{g}{c} \,\lim_{\qbar \rightarrow q}\,
                T^\mu_{\Phi g} (\qbar, \, p,\, p^\prime) \label{e16a}\\
T^\mu_{\Phi g} &= - \frac{1}{2} \frac{g^2}{2 \,\mw^2} g_s^2
\int \frac{d^n k_1}{(2 \pi)^n\, \mu^{n-4}} \frac1{k_1^2 - \mw^2}
\int \frac{d^n k_2}{(2 \pi)^n\, \mu^{n-4}} \frac1{k_2^2}
\int d^n y \,e^{-i \qbar \cdot y} \int d^n x_1 \,e^{ik_1 \cdot x_1}\nonumber \\
\times &  \int d^n x_2 \,e^{ik_2 \cdot x_2}
\int d^n x_3 \,e^{-ik_2 \cdot x_3}
 < p^\prime p \, | \, T^* \left[ J_Z^\mu (y) J_s^\lambda (x_2)
J_{s\, \lambda} (x_3)\,
( S^{\dagger}(x_1) \, S(0)+ h.c.) \right] \, | \,0 >  ,\nonumber \\
& \left. \right. \label{e16b}
\eeq \een
where $g_s$ and $J_s^\lambda$ are the $SU(3)_c$ coupling and current
respectively and we have suppressed the color indices.
Analogously to the one-loop vertex function $\cvi{\Phi}$,
$\cvi{\Phi g}$ contains not only the one-particle irreducible (1PI) two-loop
\cor\ to the proper vertex but also the contribution of the wave function
renormalization of the external lines. Concerning the latter, as shown in
Appendix A, the limiting procedure in \equ{e16a}\ does correctly take into
account the two-loop 1PI contribution of it but there is a mismatch in the
numerical coefficient of the reducible part. To cure it we replace \equ{e16a}
by \be
\widetilde{{\cal V}}_{\Phi g}^\mu =
 -i\,\frac{g}{c} \,\lim_{\qbar \rightarrow q}\,
  ( T^\mu_{\Phi g} -\pig \,T^\mu_\Phi) \label{e17}
\ee
where \be
\pig = \frac{g_s^2}{(16 \pi^2)^{n/4}} \Gamma (2 - \frac{n}2)\, (2-n)\,
       \int_0^1 d \alpha \, (1-\alpha)\, [\lambda^2(1-\alpha)]^{-2+n/2}
\label{e18} \ee
is the QCD one-loop wave function renormalization constant in $n$ dimensions
and $\lambda$ is a fictitious gluon mass introduced to regularize the IR
divergencies.
\par The four-point correlation function depicted in Fig.3(b), $\cud$,
presents
an analogous overcounting in the field renormalization of the external legs.
Analogously to \equ{e17}\ the correct factor can be obtained by
subtracting a term proportional to the one-loop two-point correlation function
$\cuu$. We write the corrected amplitude as \ben \label{e19} \beq
\cutd &= -i \frac{g}{c} \left( \cud - \pig \,\cutu \right) \label{e19a}\\
\cutd &=  \frac{i\,g^2 }{2\,\mw^2} g_s^2 (c^2- s^2)
\int \frac{d^n k_1}{(2 \pi)^n \, \mu^{n-4}} \frac{k_1^\mu}{(k_1^2-\mw^2)
[(k_1+q)^2 -\mw^2]}
\int \frac{d^n k_2}{(2 \pi)^n\, \mu^{n-4}} \frac1{k_2^2}  \nonumber \\
\times & \int d^n x_1 \,e^{ik_1 \cdot x_1} \int d^n x_2 \,e^{ik_2 \cdot x_2}
\int d^n x_3 \,e^{-ik_2 \cdot x_3}
< p^\prime p \, | \, T^* \left[  J_s^\lambda (x_2)\,
J_{s\, \lambda} (x_3)\,S^\dagger (x_1)S(0)\right] \, | \,0 >  .\nonumber\\
\left. \right. & \label{e19b}
\eeq \een
The terms $\pig\,T^\mu_\Phi$ in \equ{e17}\ and $\pig\, \cutu$ in \equ{e19a}
can be understood as follows. In the one-loop vertex functions $T^\mu_\Phi$
and $\cutu$ we can consider the bra $< p^\prime p \, |$ as dressed with
respect to the strong interactions. Expansion up to  $O(\alpha_s)$ gives
the above two contributions.
\par In order to reduce \equ{e16b}\ to a more tractable expression we apply
the same procedure developed in Section 2, namely we contract
$T^\alpha_{\Phi g} -\pig T^\alpha_\Phi$ with $\qbar_\alpha$ and then we
differentiate with respect to $\qbar_\mu$. The important point to notice
is that $J_s^\lambda$ commutes with $J_Z^\mu$. Therefore, going through
the same steps as in Section 2, we reach the two-loop counterpart of \equ{e9}\
or \beq
(T^\mu_{\Phi g}& )_{\LT} = \frac{\partial}{\partial \qbar_\mu} \dvd
 -\frac{i\,g^2 }{2\,\mw^2} g_s^2 (c^2- s^2)
\int \frac{d^n k_1}{(2 \pi)^n \, \mu^{n-4}} \frac{k_1^\mu}{(k_1^2-\mw^2)^2}
\int \frac{d^n k_2}{(2 \pi)^n\, \mu^{n-4}} \frac1{k_2^2}  \nonumber \\
\times & \int d^n x_1 \,e^{ik_1 \cdot x_1} \int d^n x_2 \,e^{ik_2 \cdot x_2}
\int d^n x_3 \,e^{-ik_2 \cdot x_3}
< p^\prime p \, | \, T^* \left[  J_s^\lambda (x_2)\,
J_{s\, \lambda} (x_3)\,S^\dagger (x_1)S(0)\right] \, | \,0 >, \nonumber \\
\left. \right. & \label{e20}
\eeq
where $\dvd$ is again an expression obtained by $T^\mu_{\Phi g}$ with the
replacement $J^\mu_Z \rightarrow -i \partial_\mu J^\mu_Z$.
The second term in the r.h.s. of \equ{e20} cancels the leading contribution
of \equ{e19b} and we are left with \be
(\widetilde{{\cal V}}_{\Phi g}^\mu + \cutd )_{\LT} =- i \frac{g}{c} \left.
\left( \frac{\partial}{\partial \qbar_\mu} \dvd - \pig
\frac{\partial}{\partial \qbar_\mu} \dvu \right) \right|_{\qbar = q}
\label{e21} \ee
\par The diagrams contributing to $\dvd$ are obtained by adding a virtual
gluon, in all possible ways, to the graphs of Fig.2. The types of diagrams
are shown in Fig.4. To every graph in the figure there corresponds
two diagrams. In the first one the momentum $k_1$ is emitted by
$S^\dagger(x)$ while in the second it is emitted by $S(x)$.
\par To get an ultraviolet (UV) finite answer we have to add to \equ{e21}
the counterterm contributions, that we indicate as $c.t.$.
The only counterterm at our disposal comes from the
bare top mass. Taking  as renormalized quantity the zero of the real
part of the inverse propagator, the so-called ``on-shell'' (OS) mass, we
have for the counterterm $\delta m_t$ \be
\delta m_t = \frac{g^2_s}{(16 \pi^2)^{n/4}} m_t \,\Gamma\, (2 -\frac{n}2) \,
      \int_0^1 d \alpha \, [2\,(1-\alpha)+n]
(m_t^2 \alpha^2)^{-2+n/2}. \label{e22} \ee
In Fig.5 the types of diagrams contributing to $c.t.$ are depicted.
The cross represents the
insertion of $\delta m_t$. Figs.5(a,b,c) are due to the presence of $m_t^0$
in the operators $S_2$ and $S$ (Cfr. \eqs{e2c}{e2e}). Similarly to Fig.4,
every graph in Fig.5 represents two contributions with different momentum
insertions.
\par Evaluation of the contributions of Fig.4 and Fig.5 (Cfr. Appendix B)
gives \be
(\widetilde{{\cal V}}_{\Phi g}^\mu + \cutd+ c.t.)_{\LT} =
 -i \frac{g}{2c} \frac{g^2}{16 \pi^2} \frac{m_t^2}{2\,\mw^2}
       \frac{g_s^2}{16 \pi^2} \left( c_1 + c_2 \xi (2) \right) C_F\,
\bar{u}(p^\prime) \gamma^\mu a_- v(p) \label{e23}
\ee
where $C_F$ is the eigenvalue of the quadratic casimir operator for
the fundamental representation of $SU(N)$, namely $(N^2-1)/(2N)$, ($C_F=4/3$
for $SU(3)_c$), \ben \label{e24} \beq
c_1 =& - \frac12 + B_r \label{e24a}\\
c_2 =& - 6 \label{e24b}
\eeq
and \be
\xi (2) = - \int_0^1 d x \frac{\ln (1-x)}{x} = \frac{\pi^2}{6}. \label{e24c}
\ee \een
In \equ{e24a}\ $B_r$ represents the contribution of the IR divergent
diagrams 4(a), 4(b), 4(c) and of $i \frac{g}{c}\left. \pig
\frac{\partial}{\partial \qbar_\mu}
\dvu  \right|_{\qbar = q}$ once the UV pole has been subtracted.
Explicitly \be
B_r = - \ln^2 \frac{\lambda^2}{q^2} - 3 \ln \frac{\lambda^2}{q^2} -3
        + \frac{\pi^2}{3}.  \label{e24d}
\ee
Comparing \equ{e23}\ with \equ{e13}\ we see that the QCD \cor\ modify
the leading $O(\alfa)$ term in the $Z b \bibar$-vertex by
$$  \frac{\alpha_s}{4 \pi} \left( c_1 + c_2 \xi (2) \right) C_F. $$
\par The inclusion of the $O(\alfas)$ \cor\ in the decay width $\Gamma (
Z \rightarrow b + \bibar)$ can be performed in the following way. We write
the $Zb\bibar$-vertex, $V^\mu$, as \be
V^\mu = -i \frac{g}{c} \gamma^\mu \frac{a+ b \gamma_5}{2} \label{e25}
\ee
where \ben \label{e26} \beq
 a=& a_0 + a_{1e}+a_{1s}+a_2 \label{e26a}\\
 b=& b_0 + b_{1e}+b_{1s}+b_2 \label{e26b}
\eeq
with \begin{eqaligntwo}
a_0=& - \frac12 + \frac23 s^2 \:;& b_0 =& \frac12 \label{e26c}\\
a_{1e}=& \frac{g^2}{16 \pi^2} \frac{m_t^2}{4\,\mw^2} \:;& b_{1e}=& - a_{1e}
\label{e26d}\\
a_{1s}=& \frac{\alpha_s}{4 \pi} a_0 c_1 C_F \:;&
b_{1s}=& \frac{\alpha_s}{4 \pi} b_0 c_1 C_F \label{e26e} \\
a_{2}=& \frac{\alpha_s}{4 \pi}a_{1e} (c_1+c_2 \xi(2))C_F \:;& b_{2}=& -
a_{2}.
\label{e26f}
\end{eqaligntwo} \een
The decay width can be written as \be
\Gamma \, (Z \rightarrow b +\bibar) = \frac{g^2}{48 \pi} \mz ( a^2 +b^2).
\label{e27}
\ee
Expanding the squares in \equ{e27}, keeping terms up to the order we are
considering, one obtains \be
\Gamma \, (Z \rightarrow b +\bibar) = \frac{g^2}{48 \pi} \mz \left[ ( a_0^2 +
b_0^2)(1+ \frac{\alpha_s}{2 \pi} c_1 C_F) + 2 (a_0- b_0)a_{1e}
\left(1+ \frac{\alpha_s}{4 \pi}(2 c_1+c_2 \xi(2)) C_F \right) \right].
\label{e28}  \ee
It is clearly understood that in \equ{e28} the factor $B_r$ present in $c_1$
is evaluated at $q^2 = \mz^2$.
\par A few remarks about the above result are now in order.
i) All the calculation
has been performed analytically. The integration over the Feynman parameters
has
been checked, always analytically, with the algebraic manipulation program
MAPLE \cite{b16}. ii) In every diagram in Fig.4 and Fig.5 but 4(a) we have
neglected all the momenta and masses compared to the top mass. In 4(a) we were
forced to keep in addition the momentum transfer $q$ due to the presence of IR
divergent terms. iii) We found it advantageous to regularize the IR
divergencies
by giving a small mass to the gluon instead of using dimensional
regularization.
In this way we did not introduce any $1/\epsilon^2$ term ($\epsilon \!
=\!n-4)$ in the calculation keeping the n-dimensional algebra simpler.
\par \equ{e28} is IR divergent in the $c_1$ coefficient (Cfr. Eqs.(\ref{e24a})
and (\ref{e24d})). As is well known, to eliminate the IR divergencies it is
necessary to take into account the QCD \bre. In particular, to be consistent
with the order of our calculation, we need to consider the \bre\ with the
inclusion of the leading $O(\alfa)$ term. The discussion of this correction
can be performed on the same footing as the $O(\alfas)$ term. In fact, the
cancellation between \equ{e19b}\ and the second term in \equ{e20}\ will work
even in the presence of a single $J_s^\lambda$ operator. The result is that
the $O(\alfa)$ term in the QCD \bre\ is obtained by adding a real
gluon in all possible ways to the diagrams in Fig.2. We note that a gluon
emitted by a virtual fermion line gives rise to a correction that does not
contribute to the leading term. It is then easy to show that the $O(\alfa)$
correction can be absorbed in a redefinition of the vector and axial coupling
of the $Z$. Therefore we can write for the rate $\Gamma (
Z \rightarrow b+\bibar + g)$
\cite{b17} \be
\Gamma (Z \rightarrow b +\bibar+ g) = \frac{g^2}{48 \pi} \mz (\tilde{a}^2
+\tilde{b}^2) \frac{\alpha_s}{2 \pi} \left[ -c_1 + \frac32 \right] C_F
\label{e29} \ee
with \ben \label{e30} \beq
\tilde{a} = & a_0 +a_{1e} \label{e30a}\\
\tilde{b} = & b_0 +b_{1e}. \label{e30b}
\eeq \een
Expanding the squares in \equ{e29}, retaining terms up to the order we are
interested,  we have\be
\Gamma (Z \rightarrow b +\bibar+ g) = \frac{g^2}{48 \pi} \mz \left(a_0^2
+b_0^2 +2(a_0 - b_0)a_{1e}\right) \frac{\alpha_s}{2 \pi}
\left[ -c_1 + \frac32 \right] C_F. \label{e31}
\ee
Finally, summing \equ{e28} and \equ{e31} one obtains \beq
\Gamma \, (Z \rightarrow b +\bibar) + \Gamma (Z \rightarrow b +\bibar+ g)
= &\frac{g^2}{48 \pi} \mz \left[ ( a_0^2 +
b_0^2)(1+ \frac{3\alpha_s}{4 \pi}  C_F) \right. \nonumber \\
& \mbox{}+ \left. 2 (a_0- b_0)a_{1e}
\left(1+ \frac{\alpha_s}{4 \pi}(3 +c_2 \xi(2)) C_F \right) \right]. \label{e32}
\eeq
\equ{e32} is the final result. The $O(\alfa)$ correction, represented
by the term $a_{1e}$ in \equ{e32}, gets modified into ($C_F= 4/3)$ \be
a_{1e} \rightarrow a_{1e} \left( 1 - \frac{\alpha_s}{\pi} \frac{\pi^2 -3}{3}
\right) . \label{e33}
\ee
Remembering that $a_{1e}$ is written as $(G_\mu m_t^2)/(\sqrt2\, 8 \pi^2)$
in the OS and $\hat{\alpha}/(4 \pi \hat{s}^2) m_t^2/(4\, \mw^2)$
in the $\overline{MS}$ formulation, we see that one can take into account
the leading QCD effect in $\rho_{bb}$ and $k_{b}$, or correspondingly
$\overline{\rho}_{bb}$ and $\hat{k}_b$, just by replacing in the
one-loop $O(\alfa)$ term $m_t^2$ by \be
m_t^2 \rightarrow m_t^2 \left( 1 - \frac{\alpha_s}{\pi} \frac{\pi^2 -3}{3}
\right) . \label{e34}
\ee
\equ{e34} coincides with the result found by Fleischer {\em et al}
\cite{b11}.
\section{Conclusions}
In the previous Sections we have shown that the use of current correlation
functions and their associated current algebra provides a very powerful and
compact framework to discuss the $O(\alfa)$ \cor\ in the $Zb\bibar$-vertex.
\par It is well known that the $O(\alfa)$ term should be finite by itself
at the one-loop level. In fact there is no counterterm available to
cancel any divergent
contribution proportional to $m_t^2$. The formalism of current correlation
functions allows one to combine several Feynman diagrams and easily enforce the
Ward identity that guarantee the finiteness of this term. The actual
calculation of the leading $O(\alfa)$ correction becomes then trivial.
\par Furthermore the current algebra framework is very suitable to discuss
the QCD \cor\ to the one-loop $O(\alfa)$ term. In fact, because the strong
and weak currents commute, the one-loop Ward identity in the \ew\ sector
is preserved. This is crucial for the structure of the divergent terms.
Indeed no poles more divergent than $1/\epsilon$ are found (Cfr. Appendix
B) because all the divergent terms are due to the QCD substructure. In
comparison the standard two-loop calculation presents
$O(1/ \epsilon^2)$, or even $O(1/ \epsilon^3)$ contributions if the
IR divergencies
are regularized using dimensional regularization \cite{b11}. From a
practical point of view this fact greatly helps in performing the calculation.
Especially the structure of the IR divergent terms is very simplified. We
find only two different IR contributions (Fig.4(a) and 4(b)) and their
evaluation in the IR part is very similar to the computation of the one-loop
QCD \cor\ to the $Z b \bibar$-vertex.
\par Concerning the physical significance of the $O(\alfas)$ calculation, we
find that the QCD \cor\ to $O(\alfa)$ term in $\Gamma (Z\rightarrow b +\bibar)$
have  opposite sign to that of the one-loop contribution. A similar situation
happens in the \ew\ parameters
$\delta \rho$ and $\Delta r$. A detailed analysis \cite{b5bis} shows
that the inclusion of the perturbative QCD higher order effects in the vacuum
polarization functions increases the prediction for $m_t$ derived from current
measurements. The values obtained for $m_t$ including the
$O(\alpha \alpha_s)$ \cor\ are larger than those obtained using only the \ew\
calculation by an amount between 5 and 10 GeV for $90 \leq m_t \leq 200$ GeV.
Using $\alpha_s=0.118$ \cite{b18}, \equ{e34} gives as correction
factor for $m_t \: 4.4 \%$, that is consistent with the result of
\efe{b5bis}. In fact,
a comparison between the correction found for $\delta \rho$, i.e\
$- \frac{2 \pi^2+6}{9} \frac{\alpha_s}{\pi} \sim -2.860 \frac{\alpha_s}{\pi}$
\cite{b4}, and  the value we obtained, $- \frac{\pi^2-3}{3}
\frac{\alpha_s}{\pi} \sim -2.290 \frac{\alpha_s}{\pi}$, shows that the two
results are numerically comparable. Both \cor\ are quite large, however, it
should be pointed out that they are scheme dependent. In our calculation, as
well as in \efe{b4}, the OS mass definition for the renormalized top mass has
been used. If we employ instead a $\overline{MS}$ definition for the top
mass,
$\hat{m}_t(\mu=\hat{m}_t)$, we have that the counterterm $\delta m_t$ has no
finite part, and therefore in the evaluation of the diagrams in Fig.5 only the
terms proportional to $\delta$ should be retained (Cfr.\ Appendix B). This
changes the coefficient multiplying $\frac{\alpha_s}{\pi}$ from
$- \frac{\pi^2-3}{3} $ to $- \frac{\pi^2-11}{3} $, i.e.\ from $\sim -2.290$
to $\sim + 0.377$, a much smaller number. Similarly the use of
$\hat{m}_t$ as renormalized mass in $\delta \rho$
changes the numerical coefficient in front of $\frac{\alpha_s}{\pi} $
from $- \frac{2 \pi^2+6}{9}$ to $- \frac{2 \pi^2-18}{9}$,  i.e.\ from
$ \sim -2.860$ to $\sim +0.193$, again a much smaller number.
\par Finally we want to stress that our calculation of $O(\alfas)$ terms
and the one presented in \efe{b11}\ are completely independent. Because in
\efe{b11}\ the IR divergencies were regularized using dimensional
regularization it is not possible to make any intermediate check, not even at
the level of $\Gamma(Z\rightarrow b +\bibar)$. It is  a welcome fact
that our result coincides with the one obtained in \efe{b11}.
\section*{Acknowledgements}
The author would like to thank A.~Sirlin for many illuminating discussions
and for informing him about \efe{b11}\ before the completion of this
project. Useful conversation with P.~Gambino are also acknowledged.
This work
was supported in part by the National Science Foundation under Grant
No.\ PHY--9017585.
\appendix
\setcounter{section}{1}
\begin{appendletter}
\section*{Appendix A}
We wish to show that Eqs.(\ref{e17}) and (\ref{e19a}) contain  correctly
the field renormalization of the external quarks. The
wave function renormalization constant $Z_b$ for a massless $b$ quark is
equal to \be
Z_b = \frac1{1 - \Pi_b} \label{ea1}
\ee
where $\Pi_b = \left. \Sigma_b ( \pisl) / \pisl \right|_{\pisl =0}$
with $\Sigma_b$ equal to $-i$
times the \self\ of the $b$. Expanding $Z_b$ up to  $O(\alpha \alpha_s)$
considering only the contributions due to the $\Phi$ and the gluon we
have \be
Z_b = 1 + \pib + \pib^2 + \cdots \simeq 1 + \pifi + \pig + \pitwo{\ }
      + 2 \,\pifi \pig + \cdots \label{ea2}
\ee
where \pifi\ is the one-loop contribution due to the $\Phi$, \pig\ the
corresponding one for the gluon and \pitwo{\ }\ the two-loop 1PI mixed
$\Phi$--gluon term. \equ{ea2} tells us that any 1PI \self\ diagram inserted
in an external leg should be multiplied by a factor 1/2 while the total
counting of the product of one-loop objects should be equal to 2.
\par We consider first the vertex function $\cvi{\Phi g}$.
Among the various contributions of
the $T^*$--product in \equ{e16b}\ there
are diagrams in which neither the strong currents nor the operators $S$ and
$S^\dagger$ enclose the vertex of the $J_Z^\mu$ current. We begin by discussing
these diagrams for the case of the $\bibar$ leg. They are depicted in
Fig.6(a--g). The wiggly line in the figure stands for the $\qbar$ momentum
absorbed by the $J_Z^\mu$ operator. Again, as in Fig.4 and Fig.5,
every graph represents two contributions with different momentum
insertions.
\par Indicating with $\pitwo{6(\mbox{$\alpha$})}$
($\alpha=$a, b, c,d) the \self\ graph, divided by the relevant momentum,
inserted in the external leg of the diagram obtained by
Fig.$6(\alpha)$ attaching together the dashed lines, we can write
the sum of the two diagrams represented by 6(a) as \be
6( {\rm a}) = \lim_{\qbar \rightarrow q}\,
-i \frac{g}{c} \bar{u}(p^\prime) \gamma^\mu \frac{a_0+b_0 \gamma_5}2
\frac{1}{\pisl^\prime - \bar{\qsl}} \frac12 \left[ \pitwo{6(a)} \pisl +
\pitwo{6(a)}(\pisl^\prime- \bar{\qsl}) \right] v(p) \label{ea3}
\ee
Remembering that \pitwo\ has the form $\pitwo{\ } = z a_+$
and observing that the first term in the square bracket in \equ{ea3}
is zero we have \be
6({\rm a}) =  -i \frac{g}{c} \frac12 z_{6({\rm a})}
\bar{u}(p^\prime) \gamma^\mu \frac{a_0+b_0 \gamma_5}2 a_- v(p) . \label{ea4}
\ee
The factor $1/2 \,z_{6({\rm a})}$
is the correct contribution of the \self\
diagram associated with Fig.6(a) to the field renormalization of the $\bibar$.
\par Let's consider now the graph 6(b). Each of the two diagrams of this
kind, called $6({\rm b}_1)$ and $6({\rm b}_2)$, can be written as \be
6({\rm b}_i)=\lim_{\qbar \rightarrow q}\, i \frac{g}{c} \bar{u}(p^\prime)
\gamma^\mu
\frac{a_0+b_0 \gamma_5}2
\frac{1}{\pisl^\prime - \bar{\qsl}} \frac12 \left[ \piov{a} \pisl + \piov{b}
(\pisl^\prime -\bar{\qsl})\right] v(p) \label{ea4b}
\ee
where $i=1,2$ and the \piovw{i}{\ } 's are related to $\pitwo{6(b)}$ through
\be
\pitwo{6(b)} =\lim_{\qbar \rightarrow q} \left( \piov{a} +\piov{b} \right) .
\label{ea5} \ee
It is clear that only the $\piov{b}$ term contributes in \equ{ea4b}.
However, an explicit calculation shows that there is a relation between
the $\piovw{i}{\ }$'s, i.e.
\begin{eqaligntwo}
\piovw{1}{a} & =  \piovw{2}{b}
; & \piovw{2}{a}& =  \piovw{1}{b}. \label{ea6}
\end{eqaligntwo}
Therefore when we sum $6({\rm b}_1)$ and
$6({\rm b}_2)$ and we use the relations
(\ref{ea6}) and \equ{ea5}\ we get \be
6({\rm b}) =  -i \frac{g}{c} \frac12 z_{6({\rm b})}
\bar{u}(p^\prime) \gamma^\mu \frac{a_0+b_0 \gamma_5}2 a_- v(p) \label{ea7}
\ee
that gives again the correct contribution to the field renormalization.
\par The sum of the diagrams 6(c) and 6(d) can be discussed in a similar
fashion as 6(b) obtaining \be
6({\rm c}) + 6({\rm d}) =
  -i \frac{g}{c} \left( \frac12 z_{6({\rm c})} +\frac12 z_{6({\rm d})}
\right)
\bar{u}(p^\prime) \gamma^\mu \frac{a_0+b_0 \gamma_5}2 a_- v(p). \label{ea7b}
\ee
Therefore we conclude that $\cvi{\Phi g}$ takes correctly into account
the two-loop 1PI contribution to the wave function renormalization.
\par Beside the 1PI contribution \equ{ea2}
contains the product of one-loop terms. Diagrams 6(e--g) represent the
corresponding contributions in $\cvi{\Phi g}$ for the $\bibar$ leg. Due to the
limiting procedure we have that each diagram carries a  factor $1/2$.
Therefore the sum (6e--g) plus the corresponding contribution of the
$b$ leg gives 3 times \pifi\ \pig. There are other
reducible contributions to $\cvi{\Phi g}$ that do not give the correct
renormalization factor. In particular diagrams 6(h) is counted  1
in $\cvi{\Phi g}$ because the limiting procedure does not affect it.
We know, however, that  it should be counted $1/2$ because of the
wave function renormalization of the external $\bibar$ leg. The same happens
for the symmetric diagram where the gluon is on the external $b$ line.
Therefore we have that in $\cvi{\Phi g}$  the
product of one-loop $\Phi$ diagrams times \pig\  is counted one time more
than the correct contribution.
The second term in \equ{e17}\ exactly cures this problem.
The vertex function $\cud$ can be discussed similarly to diagram 6(h).
The subtraction indicated in \equ{e19a}\ gives again the correct counting.
\end{appendletter}
\appendix
\setcounter{section}{2}
\begin{appendletter}
\section*{Appendix B}
In this Appendix we list the contributions of
Fig.4 to $\dvd$ and of  Fig.5 to $c.t.$. Defining
$$ \delta = \frac2{\epsilon} - 2 \gamma + 2 \ln 4 \,\pi -
2 \ln \frac{m_t^2}{\mu^2}      $$
and omitting the common factor
$$   \frac{g^2}{4}  \frac{m_t^2}{\mw^2} \frac{g_s^2}{(16\,\pi^2)^2}
C_F\,\bar{u}(p^\prime) \overline{\qsl} a_- v(p)  $$
we have \begin{eqalignno*}
4( {\rm a}) &= -\frac52   + \frac{\pi^2}{3} -\ln \frac{q^2}{m_t^2}
- \ln^2 \frac{\lambda^2}{q^2} - 4 \ln \frac{\lambda^2}{q^2} \\
4({\rm b})&=4({\rm c}) = -\delta -\frac12 + \ln \frac{\lambda^2}{m_t^2} \\
4({\rm d}) &= 4 \,\delta +13 - 6 \,\xi(2) \\
4({\rm e})&=4({\rm f}) = -\frac52 \delta -\frac{43}4 + 3 \,\xi(2) \\
4({ \rm g})&=4({\rm h}) = 4 \,\delta +\frac92 - \frac32 \xi(2) \\
4({\rm i})&=4({\rm l}) = +\frac72 - \frac32 \xi(2) \\
5({\rm a})&=5({\rm b})=5({\rm c}) = -3 \delta -7. \\
5({\rm d})&=5({\rm e}) = \frac32 \delta +\frac{13}2
\end{eqalignno*}
The left-hand-sides in the above equations indicate the appropriate
diagrams in Fig.4 and Fig.5. In the same normalization the contribution
of $ \left. - \pig\frac{\partial}{\partial \qbar_\mu} \dvu
\right|_{\qbar = q} $ is equal to $-4$(b).
\end{appendletter}

\section*{Figure Caption}
\begin{description}
\item[Fig.1] One--loop vertex diagrams contributing to the $O(\alfa)$
             corrections to the $Z\rightarrow b+ \bibar$
             process. The figures schematically represent the sum
             of Feynman diagrams in which the
             propagators are attached in all possible ways to the external
             lines.
\item[Fig.2] Momentum insertions associated with $\dvu$.
             The dotted line indicates a $\qbar$ momentum absorbed by the
             $S_2$ operator. The dashed lines stand for the momenta absorbed
             or emitted by the $S^\dagger$ and $S$ operators.
\item[Fig.3] Two-loop vertex diagrams contributing to the $O(\alfas)$
             \cor\ to the $Z\rightarrow b+ \bibar$ process.
             The meaning of the circles is explained in the text
             at the beginning of Section 3.
\item[Fig.4] Types of graphs contributing to $\dvd$. Every graph represents two
             diagrams with different momentum insertions (see text).
             The meaning of the lines is as in Fig.2.
\item[Fig.5] Types of diagrams belonging to the counterterm contributions
             $(c.t.)$. Every graph has the same meaning as in Fig.4.
\item[Fig.6]  Subset of graphs contributing to $\cvi{\Phi g}$. The meaning
             of the graphs is the same as in Fig.4 and Fig.5. The wiggly
             line stands for the $\qbar$ momentum absorbed by the $J_Z^\mu$
             operator.
\end{description}
\end{document}